\documentclass[11pt]{article}
\usepackage[a4paper]{geometry}
\usepackage{amsfonts, amsmath, amssymb, amsthm, graphicx, caption, authblk, multirow, makecell, framed, float, xcolor, mathrsfs, tikz}
\usetikzlibrary{shapes.geometric}
\DeclareMathOperator{\Ima}{Im}
\newcommand*{\mybox}[1]{%
  \framebox{\raisebox{0cm}[0.5\baselineskip][0.05\baselineskip]{%
    \hbox to 0.1cm{\hss#1\hss}}}}

\theoremstyle{definition}

\theoremstyle{remark}

\begin{document}
\tikzstyle{block} = [draw, rectangle]
\tikzstyle{line} = [draw, ->]
\tikzstyle{branch} = [draw, diamond]
\tikzstyle{finish} = [draw, circle]

\title{Securely Computing the $n$-Variable Equality Function with $2n$ Cards\thanks{A preliminary version of this paper \cite{ruangwises2} has appeared at TAMC 2020.}}
\author[1]{Suthee Ruangwises\thanks{\texttt{ruangwises@gmail.com}}}
\author[1]{Toshiya Itoh\thanks{\texttt{titoh@c.titech.ac.jp}}}
\affil[1]{Department of Mathematical and Computing Science, Tokyo Institute of Technology, Tokyo, Japan}
\date{}
\maketitle

\begin{abstract}
Research in the area of secure multi-party computation using a deck of playing cards, often called card-based cryptography, started from the introduction of the \textit{five-card trick} protocol to compute the logical AND function by den Boer in 1989. Since then, many card-based protocols to compute various functions have been developed. In this paper, we propose two new protocols that securely compute the $n$-variable \textit{equality function} (determining whether all inputs are equal) $E: \{0,1\}^n \rightarrow \{0,1\}$ using $2n$ cards. The first protocol can be generalized to compute any \textit{doubly symmetric function} $f: \{0,1\}^n \rightarrow \mathbb{Z}$ using $2n$ cards, and any symmetric function $f: \{0,1\}^n \rightarrow \mathbb{Z}$ using $2n+2$ cards. The second protocol can be generalized to compute the $k$-candidate $n$-variable equality function $E: (\mathbb{Z}/k\mathbb{Z})^n \rightarrow \{0,1\}$ using $2 \lceil \lg k \rceil n$ cards.

\textbf{Keywords:} card-based cryptography, secure multi-party computation, equality function, symmetric function, doubly symmetric function
\end{abstract}

\section{Introduction}
During a presidential election between two candidates, a group of $n$ friends wants to know whether they all support the same candidate so that they can talk about politics only if that is the case. However, each person in the group does not want to reveal his/her preference to the others (unless it is known that everyone supports the same candidate). They need a method to find out whether their preferences all coincide without leaking any other information such as the preference of any individual (not even probabilistic information).

Theoretically, this situation is equivalent to each $i$-th person having a bit $a_i$ of either 0 or 1, indicating the candidate he/she prefers. Define the \textit{equality function} $E(a_1,a_2,...,a_n) := 1$ if $a_1=a_2=...=a_n$, and $E(a_1,a_2,...,a_n) := 0$ otherwise. Our goal is to develop a protocol that announces only the value of $E(a_1,a_2,...,a_n)$ without leaking any other information.

This situation is an example of secure multi-party computation, which studies about how multiple parties can jointly compare their private information without revealing it. It is one of the most actively studied areas in cryptography. Many previous results focused on secure multi-party computation using physical objects found in everyday life such as coins \cite{coin}, combination locks \cite{diallock}, and printed transparencies \cite{darco}, but the most used object is a deck of playing cards. Hence, this research area is often called card-based cryptography. These simple protocols have an advantage that they do not require computers, and also have a great didactic value since they are easy to understand and verify the correctness and security, even for non-experts in cryptography like high school students.

\subsection{Related Work}
\subsubsection{Five-Card Trick Protocol}
The first research in card-based cryptography dates back to 1989 when den Boer \cite{denboer} developed a protocol called the \textit{five-card trick} that can compute the logical AND function of two players' bits $p$ and $q$.

The five-card trick protocol uses three identical \mybox{$\clubsuit$}s and two identical \mybox{$\heartsuit$}s, with all cards having indistinguishable back sides. At the beginning, each player is given a \mybox{$\clubsuit$} and a \mybox{$\heartsuit$}, while another \mybox{$\clubsuit$} is placed face-down on a table.

Throughout this paper, we always encode a bit 0 by a \textit{commitment} \mybox{$\clubsuit$}\mybox{$\heartsuit$} (by arranging two cards in this order), and a bit 1 by a commitment \mybox{$\heartsuit$}\mybox{$\clubsuit$}. First, the first player places a commitment of $p$ face-down to the left of the \mybox{$\clubsuit$} on the table. Then, the second player places a commitment of $q$ face-down to the right of it. Finally, publicly swap the two cards in the commitment of $q$, resulting in the following four possible sequences.

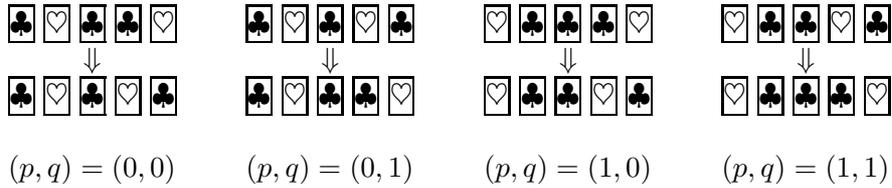
\begin{figure}[H]
    \centering
    \begin{minipage}{3cm}
        \centering
        \mybox{$\clubsuit$} \mybox{$\heartsuit$} \mybox{$\clubsuit$} \mybox{$\clubsuit$} \mybox{$\heartsuit$} \\
        $\Downarrow$ \\
        \mybox{$\clubsuit$} \mybox{$\heartsuit$} \mybox{$\clubsuit$} \mybox{$\heartsuit$} \mybox{$\clubsuit$} \\~\\
        $(p,q)=(0,0)$
    \end{minipage}
    \begin{minipage}{3cm}
        \centering
        \mybox{$\clubsuit$} \mybox{$\heartsuit$} \mybox{$\clubsuit$} \mybox{$\heartsuit$} \mybox{$\clubsuit$} \\
        $\Downarrow$ \\
        \mybox{$\clubsuit$} \mybox{$\heartsuit$} \mybox{$\clubsuit$} \mybox{$\clubsuit$} \mybox{$\heartsuit$} \\~\\
        $(p,q)=(0,1)$
    \end{minipage}
    \begin{minipage}{3cm}
        \centering
        \mybox{$\heartsuit$} \mybox{$\clubsuit$} \mybox{$\clubsuit$} \mybox{$\clubsuit$} \mybox{$\heartsuit$} \\
        $\Downarrow$ \\
        \mybox{$\heartsuit$} \mybox{$\clubsuit$} \mybox{$\clubsuit$} \mybox{$\heartsuit$} \mybox{$\clubsuit$} \\~\\
        $(p,q)=(1,0)$
    \end{minipage}
    \begin{minipage}{3cm}
        \centering
        \mybox{$\heartsuit$} \mybox{$\clubsuit$} \mybox{$\clubsuit$} \mybox{$\heartsuit$} \mybox{$\clubsuit$} \\
        $\Downarrow$ \\
        \mybox{$\heartsuit$} \mybox{$\clubsuit$} \mybox{$\clubsuit$} \mybox{$\clubsuit$} \mybox{$\heartsuit$} \\~\\
        $(p,q)=(1,1)$
    \end{minipage}
\end{figure}

Observe that there are only two possible sequences in a cyclic rotation of the sequence: \mybox{$\heartsuit$}\mybox{$\clubsuit$}\mybox{$\heartsuit$}\mybox{$\clubsuit$}\mybox{$\clubsuit$} and \mybox{$\heartsuit$}\mybox{$\heartsuit$}\mybox{$\clubsuit$}\mybox{$\clubsuit$}\mybox{$\clubsuit$}, with the latter occurring if and only if $p=q=1$. Before turning over all cards, we apply a \textit{random cut} (see Section \ref{cut}) to shift the sequence into a uniformly random cyclic shift, i.e. a permutation uniformly chosen at random from $\{\text{id}, \pi, \pi^2, \pi^3, \pi^4\}$ where $\pi = \text{(1 2 3 4 5)}$. By doing this, we can hide the initial positions of the cards and thus can determine whether $p \wedge q = 1$ without leaking any other information.

\subsubsection{Subsequent Protocols}
After the introduction of the five-card trick protocol, many other AND function protocols \cite{abe,crepeau,koch2,koch,mizuki12,mizuki09,niemi,ruangwises,stiglic} were subsequently developed. These protocols either reduced the number of required cards or improved properties of the protocol involving output format, type of shuffles, running time, etc.

Apart from the AND function protocol, protocols to compute other functions have been developed as well, such as the XOR function protocol \cite{crepeau,mizuki09,mizuki06}, the copy protocol \cite{mizuki09} (duplicating a commitment), the voting protocol \cite{mizuki13} (adding bits one by one and storing the sum in binary representation), the ranking protocol \cite{takashima} (ranking several integers without revealing them), the 3-variable \textit{majority function} protocol \cite{nishida2} (deciding whether there are more 1s than 0s in the input bits), and the 3-variable equality function protocol \cite{shinagawa}.

Nishida et al. \cite{nishida} proved that any $n$-variable Boolean function can be computed using $2n+6$ cards, and any such function that is symmetric can be computed using $2n+2$ cards.

\subsubsection{Six-Card Trick Protocol}
For the equality function, the case $n=2$ is trivial since it is simply a negation of the XOR function and thus can be computed using four cards \cite{mizuki09}. For the case $n=3$, Shinagawa and Mizuki \cite{shinagawa} developed a protocol called the \textit{six-card trick} to compute the function $E(p,q,r)$ of three players' bits $p$, $q$, and $r$ using six cards: three \mybox{$\clubsuit$}s and three \mybox{$\heartsuit$}s.

In the six-card trick protocol, each player is given a \mybox{$\clubsuit$} and a \mybox{$\heartsuit$}. First, the players place commitments of $p$, $q$, and $r$ face-down on a table in this order from left to right. Then, we rearrange the sequence of cards into a (2 4 6) permutation, resulting in the following eight possible sequences.

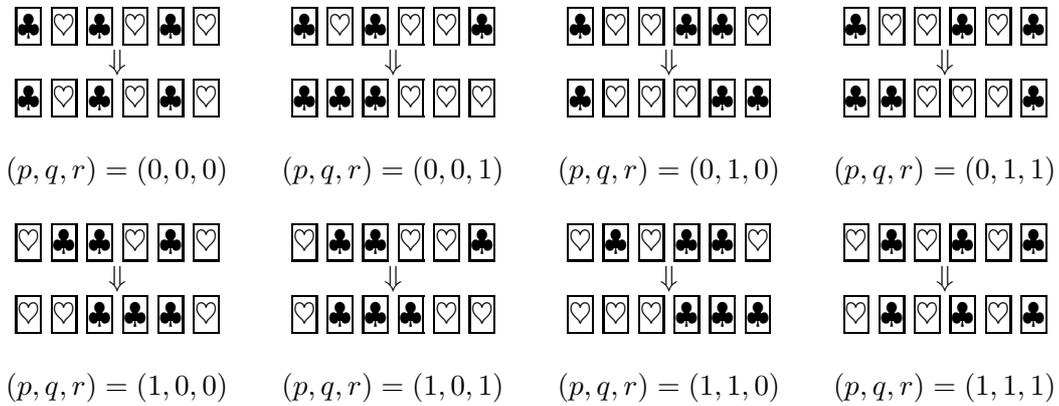
\begin{figure}[H]
    \centering
    \begin{minipage}{3.5cm}
        \centering
        \mybox{$\clubsuit$} \mybox{$\heartsuit$} \mybox{$\clubsuit$} \mybox{$\heartsuit$} \mybox{$\clubsuit$} \mybox{$\heartsuit$} \\
        $\Downarrow$ \\
        \mybox{$\clubsuit$} \mybox{$\heartsuit$} \mybox{$\clubsuit$} \mybox{$\heartsuit$} \mybox{$\clubsuit$} \mybox{$\heartsuit$} \\~\\
        $(p,q,r)=(0,0,0)$ \\~\\
				\mybox{$\heartsuit$} \mybox{$\clubsuit$} \mybox{$\clubsuit$} \mybox{$\heartsuit$} \mybox{$\clubsuit$} \mybox{$\heartsuit$} \\
        $\Downarrow$ \\
        \mybox{$\heartsuit$} \mybox{$\heartsuit$} \mybox{$\clubsuit$} \mybox{$\clubsuit$} \mybox{$\clubsuit$} \mybox{$\heartsuit$} \\~\\
        $(p,q,r)=(1,0,0)$
    \end{minipage}
    \begin{minipage}{3.5cm}
        \centering
        \mybox{$\clubsuit$} \mybox{$\heartsuit$} \mybox{$\clubsuit$} \mybox{$\heartsuit$} \mybox{$\heartsuit$} \mybox{$\clubsuit$} \\
        $\Downarrow$ \\
        \mybox{$\clubsuit$} \mybox{$\clubsuit$} \mybox{$\clubsuit$} \mybox{$\heartsuit$} \mybox{$\heartsuit$} \mybox{$\heartsuit$} \\~\\
        $(p,q,r)=(0,0,1)$ \\~\\
				\mybox{$\heartsuit$} \mybox{$\clubsuit$} \mybox{$\clubsuit$} \mybox{$\heartsuit$} \mybox{$\heartsuit$} \mybox{$\clubsuit$} \\
        $\Downarrow$ \\
        \mybox{$\heartsuit$} \mybox{$\clubsuit$} \mybox{$\clubsuit$} \mybox{$\clubsuit$} \mybox{$\heartsuit$} \mybox{$\heartsuit$} \\~\\
        $(p,q,r)=(1,0,1)$
    \end{minipage}
    \begin{minipage}{3.5cm}
        \centering
        \mybox{$\clubsuit$} \mybox{$\heartsuit$} \mybox{$\heartsuit$} \mybox{$\clubsuit$} \mybox{$\clubsuit$} \mybox{$\heartsuit$} \\
        $\Downarrow$ \\
        \mybox{$\clubsuit$} \mybox{$\heartsuit$} \mybox{$\heartsuit$} \mybox{$\heartsuit$} \mybox{$\clubsuit$} \mybox{$\clubsuit$} \\~\\
        $(p,q,r)=(0,1,0)$ \\~\\
				\mybox{$\heartsuit$} \mybox{$\clubsuit$} \mybox{$\heartsuit$} \mybox{$\clubsuit$} \mybox{$\clubsuit$} \mybox{$\heartsuit$} \\
        $\Downarrow$ \\
        \mybox{$\heartsuit$} \mybox{$\heartsuit$} \mybox{$\heartsuit$} \mybox{$\clubsuit$} \mybox{$\clubsuit$} \mybox{$\clubsuit$} \\~\\
        $(p,q,r)=(1,1,0)$
    \end{minipage}
    \begin{minipage}{3.5cm}
        \centering
        \mybox{$\clubsuit$} \mybox{$\heartsuit$} \mybox{$\heartsuit$} \mybox{$\clubsuit$} \mybox{$\heartsuit$} \mybox{$\clubsuit$} \\
        $\Downarrow$ \\
        \mybox{$\clubsuit$} \mybox{$\clubsuit$} \mybox{$\heartsuit$} \mybox{$\heartsuit$} \mybox{$\heartsuit$} \mybox{$\clubsuit$} \\~\\
        $(p,q,r)=(0,1,1)$ \\~\\
				\mybox{$\heartsuit$} \mybox{$\clubsuit$} \mybox{$\heartsuit$} \mybox{$\clubsuit$} \mybox{$\heartsuit$} \mybox{$\clubsuit$} \\
        $\Downarrow$ \\
        \mybox{$\heartsuit$} \mybox{$\clubsuit$} \mybox{$\heartsuit$} \mybox{$\clubsuit$} \mybox{$\heartsuit$} \mybox{$\clubsuit$} \\~\\
        $(p,q,r)=(1,1,1)$
    \end{minipage}
\end{figure}

Observe that there are only two possible sequences in a cyclic rotation of the sequence: \mybox{$\clubsuit$}\mybox{$\clubsuit$}\mybox{$\clubsuit$}\mybox{$\heartsuit$}\mybox{$\heartsuit$}\mybox{$\heartsuit$} and \mybox{$\clubsuit$}\mybox{$\heartsuit$}\mybox{$\clubsuit$}\mybox{$\heartsuit$}\mybox{$\clubsuit$}\mybox{$\heartsuit$}, with the latter occurring if and only if $p=q=r$. Like in the five-card trick protocol, we apply a random cut to the sequence before turning over all cards to hide the initial positions of the cards. Therefore, we can determine the value of $E(p,q,r)$ without leaking any other information.

The six-card trick protocol uses only one random cut shuffle. However, the technique used in the protocol heavily relies on the symmetric nature of the special case $n=3$, suggesting that there might not be an equivalent protocol using $2n$ cards for a general $n$. In fact, the authors \cite{shinagawa} ran a program and found that in the case $n=4$, a protocol that uses eight cards and only one random cut does not exist.

While the five-card trick and the six-card trick protocols are both useful, they have a drawback that they are not \textit{committed-format}. We call a protocol committed-format if the output is encoded in the same format as the input (here \mybox{$\clubsuit$}\mybox{$\heartsuit$} for 0 and \mybox{$\heartsuit$}\mybox{$\clubsuit$} for 1). Committed-format protocols have a benefit in the case that we do not want to reveal the output right after the computation, but instead use it as an input to another function.

\subsection{Our Contribution}
In this paper, we propose two new card-based protocols that can securely compute the $n$-variable equality function, both using $2n$ cards.

The first protocol is not committed-format, but it can be generalized to compute any \textit{doubly symmetric function} (see the definition in Section \ref{sym}) $f: \{0,1\}^n \rightarrow \mathbb{Z}$ using $2n$ cards, and any symmetric function $f: \{0,1\}^n \rightarrow \mathbb{Z}$ using $2n+2$ cards.

The second protocol is committed-format, and it can be generalized to compute the $k$-candidate $n$-variable equality function $E: (\mathbb{Z}/k\mathbb{Z})^n \rightarrow \{0,1\}$ (where there are $k$ candidates in the election instead of two) using $2 \lceil \lg k \rceil n$ cards. See Table \ref{table1} for comparison.

The major difference from the conference version of this paper \cite{ruangwises2} is the addition of our second protocol and its generalization.

\begin{table}
	\centering
	\begin{tabular}{|c|c|c|c|c|}
		\hline
		\textbf{Protocol} & \textbf{Function} & \textbf{\thead{Committed-\\ format?}} & \textbf{\#Cards} & \textbf{\#Shuffles} \\ \hline
		\thead{The six-card\\ trick \cite{shinagawa}} & \thead{equality function\\ $E: \{0,1\}^3 \rightarrow \{0,1\}$} & no & $6$ & $1$ \\ \hline
		\multirow{3}{*}{\thead{\\ \\ Nishida et al. \cite{nishida}}} & \thead{symmetric function\\ $f: \{0,1\}^n \rightarrow \{0,1\}$} & yes & $2n+2$ & $O(n \lg n)$ \\ \cline{2-5}
		& \thead{any function\\ $f: \{0,1\}^n \rightarrow \{0,1\}$} & yes & $2n+6$ & $O(n \cdot 2^n)$\\ \cline{2-5}
		& \thead{any function\\ $f: (\mathbb{Z}/k\mathbb{Z})^n \rightarrow \{0,1\}$} & yes & $2 \lceil \lg k \rceil n+6$ & $O(\lg k \cdot n(2k)^n)$ \\ \hline
		\multirow{3}{*}{\thead{\\ \\ Our first\\ protocol (\S\ref{first},\S\ref{general})}} & \thead{equality function\\ $E: \{0,1\}^n \rightarrow \{0,1\}$} & no & $2n$ & $n$ \\ \cline{2-5}
		& \thead{doubly symmetric function\\ $f: \{0,1\}^n \rightarrow \mathbb{Z}$} & no & $2n$ & $n+1+|\Ima f|$ \\ \cline{2-5}
		& \thead{symmetric function\\ $f: \{0,1\}^n \rightarrow \mathbb{Z}$} & no & $2n+2$ & $n+1+|\Ima f|$ \\ \hline
		\multirow{2}{*}{\thead{\\ Our second\\ protocol (\S\ref{second},\S\ref{general2})}} & \thead{equality function\\ $E: \{0,1\}^n \rightarrow \{0,1\}$} & yes & $2n$ & $n-1$ \\ \cline{2-5}
		& \thead{equality function\\ $E: (\mathbb{Z}/k\mathbb{Z})^n \rightarrow \{0,1\}$} & yes & $2 \lceil \lg k \rceil n$ & $\lceil \lg k \rceil n-1$ \\ \hline
	\end{tabular}
	\medskip
	\caption{Properties and number of required cards of each developed protocol} \label{table1}
\end{table}

\section{Preliminaries}
\subsection{Notations}
Given a sequence $X$ of $k$ cards, we use the following notations to denote operations on $X$.
\begin{itemize}
	\item (\texttt{perm}, $\sigma$) denotes rearranging the deck into a permutation $\sigma$, where each number $i$ represents the $i$-th leftmost card. For example, if $X=(x_1,x_2,x_3,$ $x_4,x_5,x_6)$, then (\texttt{perm}, (1 6 4)(2 5)) results in $X=(x_4,x_5,x_3,x_6,x_2,x_1)$.
	\item (\texttt{shuffle}, $\Pi$) denotes a uniformly random shuffle of $X$ unknown to all parties on the set $\Pi$ of permutations. For example, \texttt{shuffle}, \{id, (1 2 3)\}) means either doing nothing or applying (\texttt{perm}, (1 2 3)) on $X$, each  with probability 1/2.
	\item (\texttt{left\_shift}, $r$) denotes shifting $X$ to the left by $r$ positions in a cyclic manner, i.e. applying (\texttt{perm}, (1 $k$ $k-1$ ... 2)$^r$) to $X$.
	\item (\texttt{right\_shift}, $r$) denotes shifting $X$ to the right by $r$ positions in a cyclic manner, i.e. applying (\texttt{perm}, (1 2 ... $k$)$^r$) to $X$.
\end{itemize}

\subsection{Random Cut} \label{cut}
Given a sequence $X$ of $k$ cards, a \textit{random cut} or \textit{random cyclic shift} applies (\texttt{left\_shift}, $r$) to $X$, where $r$ is a uniformly random integer in $\{0,1,...,k-1\}$ unknown to all parties.

\subsubsection{Implementation}
One of the ways to implement a random cut in real world is to repeatedly apply \textit{Hindu cuts} (taking several cards from the bottom of the deck and putting them on the top) \cite{ueda}. Ideally, in each cut, observers should be able to verify that the shuffler only make a cyclic shift (and not any other permutation), but cannot tell the exact positions of the cards.\footnote{In the case where the deck has only two cards, it is obvious that every cut takes exactly one card from the bottom to the top. In this special case, we instead let all players take turn to put the deck behind their back and choose either to swap the two cards or not.} Therefore, as suggested in \cite{denboer}, if we let all players take turn to cut the cards for long enough, every player should not be able to track of the state of the deck. Moreover, the state of the deck defined by a Markov chain rapidly converges to an almost uniform distribution.

Therefore, from now on we will treat a random cut as an oracle that shifts the deck into a uniformly random cyclic shift unknown to all parties. This is a sufficient assumption for the implementation of all protocols to work correctly since all the randomness in our protocols is generated from random cuts.

\subsection{Random $k$-Section Cut}
Given a sequence $X$ of $km$ cards, a \textit{random $k$-section cut} applies (\texttt{left\_shift}, $rm$) to $X$, where $r$ is a uniformly random integer in $\{0,1,...,k-1\}$ unknown to all parties. It is equivalent to a pile-shifting shuffle developed by Shinagawa et al. \cite{shinagawa2}, and is a generalization of a \textit{random bisection cut} developed by Mizuki and Sone \cite{mizuki09}.

\subsubsection{Implementation}
The most intuitive implementation of the random $k$-section cut, as suggested in \cite{mizuki09}, is to first divide the cards into $k$ blocks, with each block consisting of $m$ consecutive cards. Then, put each block into an envelope, apply the random cut to the sequence of envelopes, and take all cards out without changing order. (Alternatively, one may instead use rubber bands or paper clips to tie the cards in each block together.)

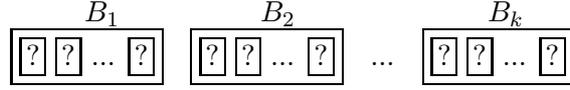
\begin{figure}[H]
    \centering
    \begin{minipage}{9cm}
				\centering
				$B_1$\hspace*{1.85cm}$B_2$\hspace*{2.53cm}$B_k$ \\
				\framebox{\mybox{?} \mybox{?} ... \mybox{?}}\hspace*{0.36cm}\framebox{\mybox{?} \mybox{?} ... \mybox{?}}\hspace*{0.36cm}...\hspace*{0.36cm}\framebox{\mybox{?} \mybox{?} ... \mybox{?}}\hspace*{0.3cm} \\~\\
		\end{minipage}
\end{figure}

Recently, Koch and Walzer \cite{koch3} proposed an alternative way to implement the random $k$-section cut without requiring extra tools by using additional cards instead.

\subsection{Random Bit XOR Protocol} \label{xor}
Given a sequence of $k$ bits $(a_1,a_2,...,a_k)$, with each $a_i$ encoded by a commitment $(x_i,y_i)$, we want to securely XOR every input bit with the same uniformly random bit $r \in \{0,1\}$ unknown to all parties, resulting in the sequence $(a_1 \oplus r, a_2 \oplus r, ..., a_k \oplus r)$. Theoretically, this is equivalent to applying (\texttt{shuffle}, \{id, (1 2)(3 4)...($2k-1$ $2k$)\}) to the sequence of cards $(x_1,y_1,x_2,y_2,...,x_n,y_n)$.

\subsubsection{Implementation}
Note that directly applying (\texttt{shuffle}, \{id, (1 2)(3 4)...($2k-1$ $2k$)\}) requires a trusted third party to first select a random bit $r$ and then apply the shuffle. Instead, the following implementation employs a random 2-section cut to achieve an equivalent result without requiring a trusted third party.

The implementation uses the same idea as the copy protocol developed by Mizuki and Sone \cite{mizuki09}. First, arrange the cards as $X = (x_1,x_2,...,x_k,y_1,y_2,...,y_k)$. Then, apply a random 2-section cut to $X$. Finally, for each $i=1,2,...,k$, take the $i$-th and the ($i+k$)-th cards from $X$ in this order as the commitment of the $i$-th output bit.

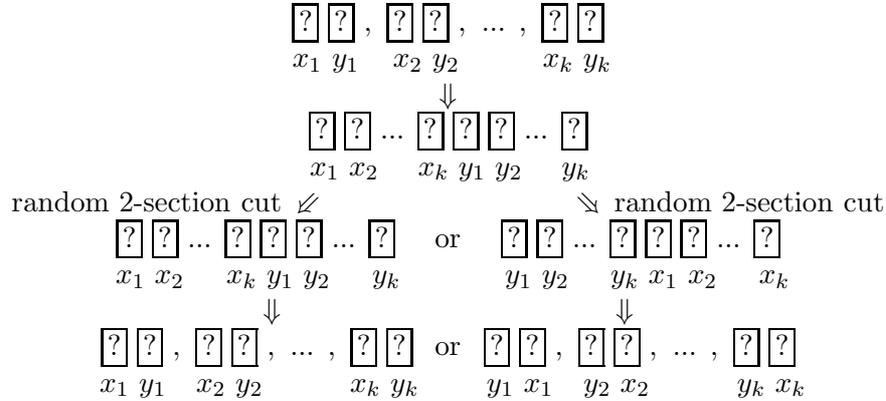
\begin{figure}[H]
		\centering
    \begin{minipage}{13cm}
				\centering
				\mybox{?} \mybox{?} ,\hspace{0.2cm}\mybox{?} \mybox{?} ,\hspace{0.2cm}... ,\hspace{0.2cm}\mybox{?} \mybox{?} \\
				\hspace{0.1cm}$x_1$ $y_1$\hspace{0.45cm}$x_2$ $y_2$\hspace{1.1cm}$x_k$ $y_k$ \\
				$\Downarrow$ \\
				\mybox{?} \mybox{?} ... \mybox{?} \mybox{?} \mybox{?} ... \mybox{?} \\
				\hspace{0.05cm}$x_1$ $x_2$\hspace{0.52cm}$x_k$ $y_1$ $y_2$\hspace{0.52cm}$y_k$ \\
				random 2-section cut \rotatebox[origin=c]{135}{$\Uparrow$}\hspace{3.3cm}\rotatebox[origin=c]{45}{$\Downarrow$} random 2-section cut \\
				\mybox{?} \mybox{?} ... \mybox{?} \mybox{?} \mybox{?} ... \mybox{?}\hspace{0.53cm}or\hspace{0.53cm}\mybox{?} \mybox{?} ... \mybox{?} \mybox{?} \mybox{?} ... \mybox{?} \\
				\hspace{0.12cm}$x_1$ $x_2$\hspace{0.55cm}$x_k$ $y_1$ $y_2$\hspace{0.55cm}$y_k$\hspace{1.38cm}$y_1$ $y_2$\hspace{0.55cm}$y_k$ $x_1$ $x_2$\hspace{0.55cm}$x_k$ \\
				$\Downarrow$\hspace{4.4cm}$\Downarrow$ \\
				\mybox{?} \mybox{?} ,\hspace{0.2cm}\mybox{?} \mybox{?} ,\hspace{0.2cm}... ,\hspace{0.2cm}\mybox{?} \mybox{?}\hspace{0.3cm}or\hspace{0.3cm}\mybox{?} \mybox{?} ,\hspace{0.2cm}\mybox{?} \mybox{?} ,\hspace{0.2cm}... ,\hspace{0.2cm}\mybox{?} \mybox{?} \\
				\hspace{0.15cm}$x_1$ $y_1$\hspace{0.4cm}$x_2$ $y_2$\hspace{1.15cm}$x_k$ $y_k$\hspace{0.9cm}$y_1$ $x_1$\hspace{0.4cm}$y_2$ $x_2$\hspace{1.15cm}$y_k$ $x_k$\hspace{0.03cm}
    \end{minipage}
\end{figure}

After applying the random 2-section cut, $X$ will become either $(x_1,x_2,...,x_k,$ $y_1,y_2,...,y_k)$ or $(y_1,y_2,...,y_k,x_1,x_2,...,x_k)$, each with probability 1/2. In the former case, the commitment of every $i$-th output bit will be $(x_i,y_i)$, which encodes $a_i \oplus 0$; in the latter case, the commitment of every $i$-th output bit will be $(y_i,x_i)$, which encodes $a_i \oplus 1$. This verifies the correctness of the protocol.

\subsection{Addition Protocol for Two Integers in $\mathbb{Z}/k\mathbb{Z}$} \label{add}
For $k \geq 3$, we introduce two encoding schemes of integers in $\mathbb{Z}/k\mathbb{Z}$ called $\clubsuit$-scheme and $\heartsuit$-scheme.

In the $\clubsuit$-scheme, an integer $a$ in $\mathbb{Z}/k\mathbb{Z}$ is encode by a sequence of $k$ consecutive face-down cards, all of them being \mybox{$\heartsuit$} except the ($a+1$)-th card from the left being \mybox{$\clubsuit$}. This sequence is called $E_k^\clubsuit(a)$, e.g. $E_3^\clubsuit(1)$ is \mybox{$\heartsuit$}\mybox{$\clubsuit$}\mybox{$\heartsuit$}, which encodes 1 in $\mathbb{Z}/3\mathbb{Z}$.

Conversely, in the $\heartsuit$-scheme, an integer $a$ in $\mathbb{Z}/k\mathbb{Z}$ is encode by a sequence of $k$ consecutive face-down cards, all of them being \mybox{$\clubsuit$} except the ($a+1$)-th card from the left being \mybox{$\heartsuit$}. This sequence is called $E_k^\heartsuit(a)$, e.g. $E_4^\heartsuit(2)$ is \mybox{$\clubsuit$}\mybox{$\clubsuit$}\mybox{$\heartsuit$}\mybox{$\clubsuit$}, which encodes 2 in $\mathbb{Z}/4\mathbb{Z}$.

Given integers $a$ and $b$ in $\mathbb{Z}/k\mathbb{Z}$, with $a$ encoded in $\heartsuit$-scheme by a sequence $X = (x_0,x_1,...,x_{k-1})$, and $b$ encoded in $\clubsuit$-scheme by a sequence $Y=(y_0,y_1,...,y_{k-1})$, we want to securely compute the sum $a+b$ (mod $k$) and have the output encoded in $\heartsuit$-scheme without using any additional card.

The idea of this protocol is that we first apply (\texttt{right\_shift}, $r$) and (\texttt{left\_ shift}, $r$) to $X$ and $Y$, respectively, for a uniformly random $r \in \mathbb{Z}/k\mathbb{Z}$ unknown to all parties, transforming $a$ and $b$ into $a-r$ and $b+r$. Then, turn over all cards in $Y$ to reveal $b+r$, and finally apply (\texttt{left\_shift}, $b+r$) to $X$ in order to make it encode $(a-r)+(b+r)=a+b$. This technique was first used by Shinagawa et al. \cite{shinagawa2} in the context of using a regular $k$-gon card to encode each integer in $\mathbb{Z}/k\mathbb{Z}$.

\subsubsection{Implementation}
Like in the random bit XOR protocol, directly applying (\texttt{right\_shift}, $r$) and (\texttt{left\_shift}, $r$) to $X$ and $Y$ with the same $r$ requires a trusted third party to first select a random integer $r$ and then apply the shifts. Instead, the following implementation employs a random $k$-section cut to achieve an equivalent result without requiring a trusted third party.

First, take the cards from $X$ and $Y$ in the following order: the leftmost card of $X$, the rightmost card of $Y$, the second leftmost card of $X$, the second rightmost card of $Y$, and so on, and place them on a single row from left to right. The cards now form a new sequence $Z = (x_0,y_{k-1},x_1,y_{k-2},...,x_{k-1},y_0)$.

\begin{figure}[H]
    \centering
		\begin{minipage}{3cm}
				\flushright
				$x_0$ $x_1$\hspace*{0.41cm}$x_{k-1}$ \\
				$X$: \mybox{?} \mybox{?} ... \mybox{?}\hspace*{0.3cm} \\
				$Y$: \mybox{?} \mybox{?} ... \mybox{?}\hspace*{0.3cm} \\
				$y_0$ $y_1$\hspace*{0.47cm}$y_{k-1}$
    \end{minipage}
		\begin{minipage}{1cm}
				\centering
				$\Rightarrow$
    \end{minipage}
		\begin{minipage}{4cm}
				\hspace*{0.55cm}$x_0$\hspace*{0.58cm}$x_1$\hspace*{0.85cm}$x_{k-1}$ \\
				$Z$: \mybox{?} \mybox{?} \mybox{?} \mybox{?} ... \mybox{?} \mybox{?} \\
				\hspace*{0.87cm}$y_{k-1}$\hspace*{0.25cm}$y_{k-2}$\hspace*{0.75cm}$y_0$
    \end{minipage}
\end{figure}

Apply a random $k$-section cut to $Z$, transforming the sequence into $(x_r,$ $y_{-r+k-1},$ $x_{r+1},y_{-r+k-2},...,x_{r+k-1},y_{-r})$ for a uniformly random $r \in \mathbb{Z}/k\mathbb{Z}$, where the indices are taken modulo $k$.

\begin{figure}[H]
    \centering
    \begin{minipage}{10.1cm}
				\hspace*{0.55cm}$x_0$\hspace*{0.55cm}$x_1$\hspace*{0.8cm}$x_{k-1}$\hspace*{2.34cm}$x_r$\hspace*{0.72cm}$x_{r+1}$\hspace*{1.06cm}$x_{r+k-1}$ \\
				$Z$: \mybox{?} \mybox{?} \mybox{?} \mybox{?} ... \mybox{?} \mybox{?}\hspace*{0.7cm}$\Rightarrow$\hspace*{0.5cm}$Z$: \mybox{?}\hspace*{0.3cm}\mybox{?}\hspace*{0.3cm}\mybox{?}\hspace*{0.3cm}\mybox{?}\hspace*{0.3cm}...\hspace*{0.3cm}\mybox{?}\hspace*{0.3cm}\mybox{?} \\
				\hspace*{0.8cm}$y_{k-1}$\hspace*{0.25cm}$y_{k-2}$\hspace*{0.83cm}$y_0$\hspace*{2.34cm}$y_{-r+k-1}$\hspace*{0.2cm}$y_{-r+k-2}$\hspace*{0.78cm}$y_{-r}$
    \end{minipage}
\end{figure}

Take the cards in $Z$ from left to right and place them at these positions in $X$ and $Y$ in the following order: the leftmost position of $X$, the rightmost position of $Y$, the second leftmost position of $X$, the second rightmost position of $Y$, and so on. We now have sequences $X = (x_r,x_{r+1},...,x_{r+k-1})$ and $Y = (y_{-r},y_{-r+1},...,y_{-r+k-1})$, which encode $a-r$ (mod $k$) and $b+r$ (mod $k$), respectively.

\begin{figure}[H]
    \centering
		\begin{minipage}{5cm}
				\hspace*{0.55cm}$x_r$\hspace*{0.72cm}$x_{r+1}$\hspace*{1.06cm}$x_{r+k-1}$ \\
				$Z$: \mybox{?}\hspace*{0.3cm}\mybox{?}\hspace*{0.3cm}\mybox{?}\hspace*{0.3cm}\mybox{?}\hspace*{0.3cm}...\hspace*{0.3cm}\mybox{?}\hspace*{0.3cm}\mybox{?} \\
				\hspace*{0.84cm}$y_{-r+k-1}$\hspace*{0.2cm}$y_{-r+k-2}$\hspace*{0.75cm}$y_{-r}$
    \end{minipage}
		\begin{minipage}{0.7cm}
				\centering
				$\Rightarrow$
    \end{minipage}
		\begin{minipage}{4cm}
				\flushright
				$x_r$\hspace*{0.23cm}$x_{r+1}$\hspace*{0.5cm}$x_{r+k-1}$\hspace*{0.16cm} \\
				$X:$ \mybox{?}\hspace*{0.3cm}\mybox{?}\hspace*{0.3cm}...\hspace*{0.3cm}\mybox{?}\hspace*{0.88cm} \\
				$Y:$ \mybox{?}\hspace*{0.3cm}\mybox{?}\hspace*{0.3cm}...\hspace*{0.3cm}\mybox{?}\hspace*{0.88cm} \\
				$y_{-r}$ $y_{-r+1}$\hspace*{0.3cm}$y_{-r+k-1}$
    \end{minipage}
\end{figure}

Turn over all cards in $Y$ to reveal $s = b+r$. Note that this revelation does not leak any information about $b$ because $b+r$ has an equal probability to be any integer in $\mathbb{Z}/k\mathbb{Z}$ no matter what $b$ is (see \ref{app1.1} for a formal proof). Finally, we apply (\texttt{left\_shift}, $s$) to $X$, transforming it into $(x_{r-s},x_{r-s+1},...,x_{r-s+k-1})$.

\begin{figure}[H]
    \centering
		\begin{minipage}{8.5cm}
				$X:$ \mybox{?}\hspace*{0.17cm}\mybox{?}\hspace*{0.17cm}...\hspace*{0.17cm}\mybox{?}\hspace*{0.7cm}$\Rightarrow$\hspace*{0.5cm}$X:$ \mybox{?}\hspace*{0.4cm}\mybox{?}\hspace*{0.4cm}...\hspace*{0.4cm}\mybox{?} \\
				\hspace*{0.64cm}$x_r$ $x_{r+1}$ $x_{r+k-1}$\hspace*{1.52cm}$x_{r-s}$ $x_{r-s+1}$\hspace*{0.15cm}$x_{r-s+k-1}$
    \end{minipage}
\end{figure}

Therefore, we now have a sequence $X$ encoding $a-r+s \equiv (a-r)+(b+r) \equiv a+b$ (mod $k$) in $\heartsuit$-scheme as desired.

\section{Our First Protocol} \label{first}
We get back to the main problem of computing the equality function $E(a_1,$ $a_2,...,a_n)$. For each $k=1,2,...,n$, let $s_k = \sum_{i=1}^k a_i$ (the sum in $\mathbb{Z}$, not in $\mathbb{Z}/2\mathbb{Z}$). Observe that the value of $E(a_1,a_2,...,a_n)$ depends only on $s_n$. Therefore, we will instead develop a protocol to compute $s_n$.\footnote{Although there is a voting protocol of Mizuki et al. \cite{mizuki13} that can compute the sum of $n$ input bits, the output of their protocol are stored in binary representation, which is not a usable format for the final step of our protocol to hide the actual sum.} The idea of this protocol is that for each $k=2,3,...,n$, we inductively compute $s_k$ in $\mathbb{Z}/(k+1)\mathbb{Z}$. Since $s_k$ is at most $k$, its value in $\mathbb{Z}/(k+1)\mathbb{Z}$ does not change from its actual value in $\mathbb{Z}$.

\subsection{Summation of the First $k$ Bits} \label{sum}
We will show that given two additional cards, one \mybox{$\clubsuit$} and one \mybox{$\heartsuit$}, we can compute $s_k$ for every $k=2,3,...,n$.

First, swap the two cards in the commitment of $a_1$ and place an additional \mybox{$\clubsuit$} face-down to the right of them. The resulting sequence, called $C_1$, is $E_3^\heartsuit(a_1)$.

\begin{figure}[H]
    \centering
    \begin{minipage}{3.8cm}
				\begin{flushright}
        Case $a_1=0$: \hspace*{0.6cm}$\clubsuit$\hspace*{0.2cm}$\heartsuit$\hspace*{0.04cm} \\
				Case $a_1=1$: \hspace*{0.6cm}$\heartsuit$\hspace*{0.2cm}$\clubsuit$\hspace*{0.04cm} \\
				$a_1$: \mybox{?} \mybox{?}
				\end{flushright}
    \end{minipage}
		\begin{minipage}{0.67cm}
        \centering
        \centering
        ~\\~\\
				$\Rightarrow$
    \end{minipage}
		\begin{minipage}{0.94cm}
        \centering
        $\heartsuit$\hspace*{0.2cm}$\clubsuit$ \\
				$\clubsuit$\hspace*{0.2cm}$\heartsuit$ \\
				\mybox{?} \mybox{?}
    \end{minipage}
		\begin{minipage}{0.67cm}
        \centering
        ~\\~\\
				$\Rightarrow$
    \end{minipage}
		\begin{minipage}{2.01cm}
        \hspace*{0.7cm}$\heartsuit$\hspace*{0.2cm}$\clubsuit$\hspace*{0.2cm}$\clubsuit$ \\
				\hspace*{0.7cm}$\clubsuit$\hspace*{0.2cm}$\heartsuit$\hspace*{0.2cm}$\clubsuit$ \\
				$C_1$: \mybox{?} \mybox{?} \mybox{$\clubsuit$}
    \end{minipage}
\end{figure}

Then, place an additional \mybox{$\heartsuit$} face-down to the right of the commitment of $a_2$. The resulting sequence, called $C_2$, is $E_3^\clubsuit(a_2)$.

\begin{figure}[H]
    \centering
    \begin{minipage}{3.8cm}
				\begin{flushright}
        Case $a_2=0$: \hspace*{0.6cm}$\clubsuit$\hspace*{0.2cm}$\heartsuit$\hspace*{0.04cm} \\
				Case $a_2=1$: \hspace*{0.6cm}$\heartsuit$\hspace*{0.2cm}$\clubsuit$\hspace*{0.04cm} \\
				$a_2$: \mybox{?} \mybox{?}
				\end{flushright}
    \end{minipage}
		\begin{minipage}{0.67cm}
        \centering
        ~\\~\\
				$\Rightarrow$
    \end{minipage}
		\begin{minipage}{2.01cm}
        \hspace*{0.7cm}$\clubsuit$\hspace*{0.2cm}$\heartsuit$\hspace*{0.2cm}$\heartsuit$ \\
				\hspace*{0.7cm}$\heartsuit$\hspace*{0.2cm}$\clubsuit$\hspace*{0.2cm}$\heartsuit$ \\
				$C_2$: \mybox{?} \mybox{?} \mybox{$\heartsuit$}
    \end{minipage}
\end{figure}

Apply the addition protocol in Section \ref{add} to compute the sum $a_1+a_2 = s_2$ and have the output stored in $C_1$, which is now $E_3^\heartsuit(s_2)$. We also have one \mybox{$\clubsuit$} and two \mybox{$\heartsuit$}s from $C_2$ after we turned them over. These cards are called \textit{free cards} and are available for us to use later in this protocol.

\begin{figure}[H]
    \centering
    \begin{minipage}{3.19cm}
				\begin{flushright}
        $C_1$ encoding $s_2$: \\~\\
				free cards from $C_2$:
				\end{flushright}
    \end{minipage}
		\begin{minipage}{2.53cm}
        \mybox{?} \mybox{?} \mybox{?} \\~\\
				\mybox{$\clubsuit$} \mybox{$\heartsuit$} \mybox{$\heartsuit$}
    \end{minipage}
\end{figure}

Inductively, for each $k=3,4,...,n$, after we finish computing $s_{k-1}$, we now have a sequence $C_1$ being $E_k^\heartsuit(s_{k-1})$. We also have one free \mybox{$\clubsuit$} and $k-1$ free \mybox{$\heartsuit$}s from $C_{k-1}$ after we turned them over. Append the free \mybox{$\clubsuit$} face-down to the right of $C_1$, making the sequence become $E_{k+1}^\heartsuit(s_{k-1})$. Also, place the $k-1$ free \mybox{$\heartsuit$}s face-down to the right of the commitment of $a_k$. The resulting sequence, called $C_k$, is now $E_{k+1}^\clubsuit(a_k)$.

\begin{figure}[H]
    \centering
    \begin{minipage}{3.41cm}
				\begin{flushright}
        $C_1$ encoding $s_{k-1}$: \\~\\
				commitment of $a_k$:
				\end{flushright}
    \end{minipage}
		\begin{minipage}{2.53cm}
        \mybox{?} \mybox{?} ... \mybox{?} \\~\\
				\mybox{?} \mybox{?}
    \end{minipage}
		\begin{minipage}{0.67cm}
        \centering
        $\Rightarrow$ \\~\\
				$\Rightarrow$
    \end{minipage}
		\begin{minipage}{3.2cm}
				\begin{flushright}
        $C_1$ encoding $s_{k-1}$: \\~\\
				$C_k$ encoding $a_k$:
				\end{flushright}
    \end{minipage}
		\begin{minipage}{3.03cm}
        \mybox{?} \mybox{?} ... \mybox{?} \mybox{$\clubsuit$} \\~\\
				\mybox{?} \mybox{?} \mybox{$\heartsuit$} ... \mybox{$\heartsuit$}
    \end{minipage}
		\vspace{0.3cm}
		
		\begin{minipage}{3.6cm}
				\begin{flushright}
				free cards from $C_{k-1}$:
				\end{flushright}
    \end{minipage}
		\begin{minipage}{3cm}
				$1 \times$\mybox{$\clubsuit$} , $(k-1) \times$\mybox{$\heartsuit$}
    \end{minipage}
		\begin{minipage}{7cm}
		~
    \end{minipage}
\end{figure}

Then, apply the addition protocol to compute the sum $s_{k-1}+a_k = s_k$ and have the output stored in $C_1$, which is now $E_{k+1}^\heartsuit(s_k)$ as desired.

\begin{figure}[H]
    \centering
    \begin{minipage}{3.2cm}
				\begin{flushright}
        $C_1$ encoding $s_k$: \\~\\
				free cards from $C_k$:
				\end{flushright}
    \end{minipage}
		\begin{minipage}{2.5cm}
        \mybox{?} \mybox{?} ... \mybox{?} \\~\\
				$1 \times$\mybox{$\clubsuit$} , $k \times$\mybox{$\heartsuit$}
    \end{minipage}
\end{figure}

Therefore, starting with one additional \mybox{$\clubsuit$} and one additional \mybox{$\heartsuit$}, we can compute the sum $s_k = \sum_{i=1}^k a_i$ for every $k=2,3,...,n$.

For example, if $a_1=1$, $a_2=0$, and $a_3=1$, at first we will have $C_1 = E_3^\heartsuit(1)$ encoding $a_1$ and $C_2 = E_3^\clubsuit(0)$ encoding $a_2$. After applying the addition protocol for the first time, we will have $C_1 = E_3^\heartsuit(1)$ encoding $s_2$. After appending free cards, we will have $C_1 = E_4^\heartsuit(1)$ encoding $s_2$ and $C_2 = E_4^\clubsuit(1)$ encoding $a_3$. After applying the addition protocol for the second time, we will have $C_1 = E_4^\heartsuit(2)$ encoding $s_3$ as desired.

\subsection{The Complete Protocol} \label{firstreal}
The protocol in Section \ref{sum} requires two additional cards to compute any $s_k$. However, we can compute the equality function without using any additional card by the following protocol.

First, apply the random bit XOR protocol in Section \ref{xor} to transform the input sequence into $(a'_1,a'_2,...,a'_n) = (a_1 \oplus r, a_2 \oplus r, ..., a_n \oplus r)$ for a uniformly random bit $r \in \{0,1\}$. Then, turn over the two cards encoding $a'_n$ to reveal $a_n \oplus r$. Note that this revelation does not leak any information about $a_n$ because $a_n \oplus r$ has an equal probability to be 0 or 1 no matter what $a_n$ is (see \ref{app1.2} for a formal proof).

If $a_n \oplus r = 0$, then $E(a_1,a_2,...,a_n) = 1$ if and only if $a_i \oplus r = 0$ for every $i=1,2,...,n-1$, which is equivalent to $\sum_{i=1}^{n-1} a'_i = 0$. Note that we now have one free \mybox{$\clubsuit$} and one free \mybox{$\heartsuit$} from the cards encoding $a'_n$ we just turned over. With these two additional cards, we can apply the protocol in Section \ref{sum} to compute the sum $\sum_{i=1}^{n-1} a'_i$ as desired.

If $a_n\oplus r = 1$, then $E(a_1,a_2,...,a_n) = 1$ if and only if $a_i \oplus r = 1$ for every $i=1,2,...,n-1$, which is equivalent to $\sum_{i=1}^{n-1} \overline{a'_i} = 0$. Therefore, we can swap the two cards encoding every bit so that each $i$-th bit becomes $\overline{a'_i}$, and apply the same protocol to compute the sum $\sum_{i=1}^{n-1} \overline{a'_i}$.

Note that the final sum $s$ of $n-1$ bits is encoded in $\heartsuit$-scheme by a row of $n$ cards, and we have $E(a_1,a_2,...,a_n) = 1$ if and only if $s$ is zero, i.e. the \mybox{$\heartsuit$} card is located at the leftmost position. However, we do not want to reveal any information about $s$ except whether it is zero or not. Therefore, we apply a final random cut to the sequence of $n-1$ rightmost cards (all cards in the row except the leftmost one) to make all the cases where $s$ is not zero indistinguishable (see \ref{app1.3} for a formal proof). Finally, turn over all cards and locate the position of the \mybox{$\heartsuit$}. If it is at the leftmost position, then $E(a_1,a_2,...,a_n)=1$; otherwise, $E(a_1,a_2,...,a_n)=0$.

Throughout this protocol, we use one random 2-section cut in the random bit XOR protocol, $n-2$ random $k$-section cuts to compute the sum of $n-1$ bits, and one final random cut. Therefore, this protocol uses $2n$ cards and $n$ shuffles.

\section{Generalization of Our First Protocol} \label{general}
\subsection{Computing Other Symmetric Functions} \label{sym}
A function $f: \{0,1\}^n \rightarrow \mathbb{Z}$ is symmetric if
$$f(a_1,a_2,...,a_n) = f(a_{\sigma_1},a_{\sigma_2},...,a_{\sigma_n})$$
for any $a_1,a_2,...,a_n$ and any permutation $(\sigma_1,\sigma_2,...,\sigma_n)$ of $(1,2,...,n)$. A symmetric function $f$ is doubly symmetric if
$$f(a_1,a_2,...,a_n) = f(\overline{a_1},\overline{a_2},...,\overline{a_n})$$
for any $a_1,a_2,...,a_n$. For example, the equality function is doubly symmetric, while the majority function is symmetric but not doubly symmetric. Other examples of doubly symmetric functions include $f_1(a_1,a_2,...,a_n) := a_1 \oplus a_2 \oplus ... \oplus a_n$ for an even $n$, and a function $f_2$ defined to be the difference of the number of 0s and 1s in the input bits.

For any symmetric function $f: \{0,1\}^n \rightarrow \mathbb{Z}$, observe that any two $n$-tuples of input bits with the same number of 1s always yield the same output (because we can rearrange the order of bits in one $n$-tuple to become another one). Hence, the value of $f(a_1,a_2,...,a_n)$ depends only on the sum $\sum_{i=1}^n a_i$ (the sum in $\mathbb{Z}$), so $f$ can be written as
$$f(a_1,a_2,...,a_n) = g\left(\sum_{i=1}^n a_i\right)$$
for some function $g:\{0,1,...,n\} \rightarrow \mathbb{Z}$. Also, if $f$ is doubly symmetric, we have $g(a) = g(n-a)$ for every $a \in \{0,1,...,n\}$.

The technique used in our first protocol can be applied to compute any doubly symmetric function. Let $f: \{0,1\}^n \rightarrow \mathbb{Z}$ be any doubly symmetric function, and let $g:\{0,1,...,n\} \rightarrow \mathbb{Z}$ be a function such that
$$f(a_1,a_2,...,a_n) = g\left(\sum_{i=1}^n a_i\right).$$
Like in our first protocol, we apply the random bit XOR protocol in Section \ref{xor} to transform the input sequence into $(a'_1,a'_2,...,a'_n) = (a_1 \oplus r, a_2 \oplus r, ..., a_n \oplus r)$ for a uniformly random bit $r \in \{0,1\}$. Then, turn over the two cards encoding $a'_n$ to reveal $a_n \oplus r$. This does not leak any information about $a_n$ because $a_n \oplus r$ has an equal probability to be 0 or 1 no matter what $a_n$ is (see \ref{app1.2} for a formal proof).

Since $f$ is doubly symmetric, if $a_n \oplus r = 0$, then
\begin{align*}
f(a_1,a_2,...,a_n) &= f(a_1 \oplus r, a_2 \oplus r, ..., a_n \oplus r) \\
&= g\left(\sum_{i=1}^n (a_i \oplus r)\right) \\
&= g\left(\sum_{i=1}^{n-1} (a_i \oplus r)\right) \\
&= g\left(\sum_{i=1}^{n-1} a'_i\right).
\end{align*}
Therefore, we can apply the protocol in Section \ref{sum} to compute the sum $\sum_{i=1}^{n-1} a'_i$ as desired.

If $a_n \oplus r = 1$, then we have $\overline{a_n \oplus r} = 0$ and
\begin{align*}
f(a_1,a_2,...,a_n) &= f(\overline{a_1 \oplus r}, \overline{a_2 \oplus r}, ..., \overline{a_n \oplus r}) \\
&= g\left(\sum_{i=1}^n \overline{a_i \oplus r}\right) \\
&= g\left(\sum_{i=1}^{n-1} \overline{a_i \oplus r}\right) \\
&= g\left(\sum_{i=1}^{n-1} \overline{a'_i}\right).
\end{align*}
Therefore, we can swap the two cards encoding every bit so that each $i$-th bit becomes $\overline{a'_i}$, and apply the same protocol to compute the sum $\sum_{i=1}^{n-1} \overline{a'_i}$.

For each $b \in \Ima f = \Ima g$, let $P_b = \{a \in \{0,1,...,n-1\} | g(a)=b\}$. Like in our first protocol, the final sum $s$ of $n-1$ bits is encoded in $\heartsuit$-scheme by a row of $n$ cards. Recall that in $\heartsuit$-scheme, $s$ is encoded by $E_n^\heartsuit(s)$ where the only \mybox{$\heartsuit$} is located at the ($s+1$)-th leftmost position. Therefore, we take the ($a+1$)-th leftmost cards from the row for all $a \in P_b$, apply a random cut to them, and put them back into the row at their original positions. We perform this in order to make all the cases where the sum is in $P_b$ indistinguishable. For example, if $P_b = \{1,4,6\}$, we take the 2nd, 5th, and 7th leftmost cards from the row, apply a random cut to the sequence of these three cards, and put them back at the 2nd, 5th, and 7th leftmost position in the row (where the original 2nd card may now end up at the 2nd, 5th, or 7th position with equal probability).

We separately apply the above random cut for every $b \in \Ima f$ such that $|P_b| > 1$. These random cuts ensure that turning over the cards does not reveal any information about $s$ except the value of $g(s)$ (see \ref{app1.3} for a formal proof). Finally, we turn over all cards. Suppose the shown sequence is $E_n^\heartsuit(t)$ for some integer $t$, then $f(a_1,a_2,...,a_n) = g(t)$.

Throughout this protocol, we use one random 2-section cut in the random bit XOR protocol, $n-2$ random $k$-section cuts to compute the sum of $n-1$ bits, and at most $|\Ima f|$ final random cuts. Therefore, this protocol uses $2n$ cards and at most $n-1+|\Ima f|$ shuffles.

For a function $f$ that is symmetric but not doubly symmetric, we directly apply the protocol in Section \ref{sum} to compute the sum $s_n = \sum_{i=1}^n a_i$ and apply the final random cuts the same way as above. It requires two additional cards, one \mybox{$\clubsuit$} and one \mybox{$\heartsuit$}, at the beginning. Therefore, this protocol uses $2n+2$ cards and at most $n-1+|\Ima f|$ shuffles.

\subsection{Comparison with Other Protocols}
There is a voting protocol developed by Mizuki et al. \cite{mizuki13} that can compute the sum of $n$ input bits using only $O(\log n)$ cards. However, their protocol restricts the order of submission of the inputs in order to reuse cards encoding each input bit. Any protocol that the inputs are submitted simultaneously requires at least $2n$ cards as we need at least two cards to encode each input bit. Hence, our protocol is the optimal one for computing any doubly symmetric function, improving the previous protocol of Nishida et al. \cite{nishida} which requires $2n+2$ cards.

For computing symmetric functions that are not doubly symmetric, the protocol of \cite{nishida} also uses $2n+2$ cards to compute any symmetric function $f: \{0,1\}^n \rightarrow \{0,1\}$, but it uses as much as $O(n \lg n)$ shuffles due to the construction of every term of Shannon expansion. Their protocol has an advantage that it is committed-format. However, our protocol uses much fewer number of shuffles and also has an advantage that the output is not restricted to be binary, thus supporting functions with more than two possible outputs. (An example of such function is the majority function that supports the case of a tie for an even $n$, which has three possible outputs.)

\section{Our Second Protocol} \label{second}
In this section, we will introduce a second protocol to compute the equality function $E(a_1,a_2,...,a_n)$.

\subsection{AND function protocol} \label{and}
First, we will explain a protocol developed by Mizuki and Sone \cite{mizuki09} that can compute the AND function of two bits using two additional cards, one \mybox{$\clubsuit$} and one \mybox{$\heartsuit$}.

Suppose we have commitments $(x_1,y_1)$ and $(x_2,y_2)$ of bits $a$ and $b$, respectively. We place the two additional cards face-down as \mybox{$\clubsuit$}\mybox{$\heartsuit$}, called $(x_3,y_3)$, to form a commitment of 0.

Arrange the cards as $X = (x_1,y_1,x_2,y_2,x_3,y_3)$, which encodes a sequence of bits $(a,b,0)$. Then, apply (\texttt{perm}, (2 4 3)), apply a random 2-section cut, and apply (\texttt{perm}, (2 4 3)) to $X$. Note that these steps are theoretically equivalent to applying (\texttt{shuffle}, \{id, (1 2)(3 5)(4 6)\}) to $X$.

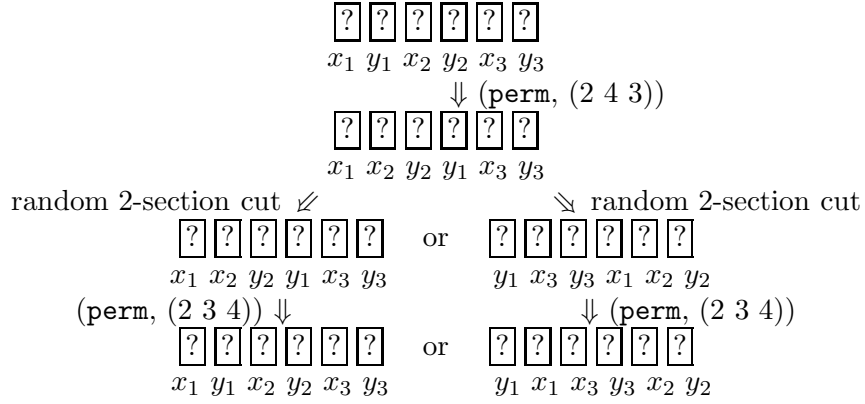
\begin{figure}[H]
		\centering
    \begin{minipage}{13cm}
				\centering
				\mybox{?} \mybox{?} \mybox{?} \mybox{?} \mybox{?} \mybox{?} \\
				$x_1$ $y_1$ $x_2$ $y_2$ $x_3$ $y_3$ \\
				\hspace*{3.2cm}$\Downarrow$ (\texttt{perm}, (2 4 3)) \\
				\mybox{?} \mybox{?} \mybox{?} \mybox{?} \mybox{?} \mybox{?} \\
				$x_1$ $x_2$ $y_2$ $y_1$ $x_3$ $y_3$ \\
				random 2-section cut \rotatebox[origin=c]{135}{$\Uparrow$}\hspace{3cm}\rotatebox[origin=c]{45}{$\Downarrow$} random 2-section cut \\
				\mybox{?} \mybox{?} \mybox{?} \mybox{?} \mybox{?} \mybox{?}\hspace{0.53cm}or\hspace{0.53cm}\mybox{?} \mybox{?} \mybox{?} \mybox{?} \mybox{?} \mybox{?} \\
				\hspace{0.12cm}$x_1$ $x_2$ $y_2$ $y_1$ $x_3$ $y_3$\hspace{1.38cm}$y_1$ $x_3$ $y_3$ $x_1$ $x_2$ $y_2$ \\
				(\texttt{perm}, (2 3 4)) $\Downarrow$\hspace{3.8cm}$\Downarrow$ (\texttt{perm}, (2 3 4)) \\
				\mybox{?} \mybox{?} \mybox{?} \mybox{?} \mybox{?} \mybox{?}\hspace{0.53cm}or\hspace{0.53cm}\mybox{?} \mybox{?} \mybox{?} \mybox{?} \mybox{?} \mybox{?} \\
				\hspace{0.15cm}$x_1$ $y_1$ $x_2$ $y_2$ $x_3$ $y_3$\hspace{1.38cm}$y_1$ $x_1$ $x_3$ $y_3$ $x_2$ $y_2$
    \end{minipage}
\end{figure}

After these operations, $X$ now becomes either $(x_1,y_1,x_2,y_2,x_3,y_3)$ or $(y_1,x_1,$ $x_3,y_3,x_2,y_2)$, each with probability 1/2. The former encodes $(a,b,0)$, and the latter encodes $(\overline{a},0,b)$.

We turn over the two leftmost cards to reveal the first bit. This does not leak any information about $a$ because the first bit has an equal probability to be 0 or 1 no matter what $a$ is. If the first bit is 0, then the third bit is the output of the AND function; if the first bit is 1, then the second bit is the output of the AND function. This is true because
$$
a \wedge b = \begin{cases}
0, &\text{if } a=0; \\
b, &\text{if } a=1. \\
\end{cases}
$$

This protocol is committed-format, so we can use the output as an input to another function. Therefore, we can compute $a_1 \wedge a_2 \wedge ... \wedge a_k$ by computing $(((a_1 \wedge a_2) \wedge a_3) \wedge ...) \wedge a_k$ in this order. Moreover, the two additional cards can be reused in each computation, so only two additional cards are sufficient for any $k$.

\subsection{The Complete Protocol}
To compute the equality function, we start in the same way as our first protocol by applying the random bit XOR protocol to transform the input sequence into $(a'_1,a'_2,...,a'_n) = (a_1 \oplus r, a_2 \oplus r, ..., a_n \oplus r)$ for a uniformly random bit $r \in \{0,1\}$. Then, turn over the two cards encoding $a'_n$ to reveal $a_n \oplus r$.

If $a_n \oplus r = 1$, then $E(a_1,a_2,...,a_n) = 1$ if and only if $a_i \oplus r = 1$ for every $i=1,2,...,n-1$, which is equivalent to $a'_1 \wedge a'_2 \wedge ... \wedge a'_{n-1} = 1$. Therefore, $E(a_1,a_2,...,a_n) = a'_1 \wedge a'_2 \wedge ... \wedge a'_{n-1}$, which can be computed by applying the AND function protocol in section \ref{and} using two free cards from the cards encoding $a'_n$ we just turned over.

If $a_n \oplus r = 0$, then $E(a_1,a_2,...,a_n) = 1$ if and only if $a_i \oplus r = 0$ for every $i=1,2,...,n-1$, which is equivalent to $\overline{a'_1} \wedge \overline{a'_2} \wedge ... \wedge \overline{a'_{n-1}}= 1$. Therefore, $E(a_1,a_2,...,a_n) = \overline{a'_1} \wedge \overline{a'_2} \wedge ... \wedge \overline{a'_{n-1}}$, which can be computed by swapping the two cards encoding each bit and applying the same protocol.

Throughout this protocol, we use one random 2-section cut in the random bit XOR protocol and $n-2$ random 2-section cuts to compute the AND function of $n-1$ bits. Therefore, this protocol uses $2n$ cards and $n-1$ shuffles, one less shuffle than our first protocol, and also has an advantage that it is committed-format.

\section{Generalization of Our Second Protocol} \label{general2}
The technique used in our second protocol relies on the unique property of the equality function, so it cannot be applied to compute other symmetric functions. However, we can use a similar technique to compute the $k$-candidate $n$-variable equality function $E: (\mathbb{Z}/k\mathbb{Z})^n \rightarrow \{0,1\}$ (where there are $k$ candidates in the election instead of two) using $2 \lceil \lg k \rceil n$ cards for any $k \geq 3$.

\subsection{Computing the $k$-Candidate $n$-Variable Equality Function}
Analogously to the Boolean equality function, for $a_1,a_2,...,a_n \in \mathbb{Z}/k\mathbb{Z}$, define $E(a_1,a_2,...,a_n)$ $:= 1$ if $a_1=a_2=...=a_n$, and $E(a_1,a_2,...,a_n) := 0$ otherwise.

Let $\ell = \lceil \lg k \rceil$. Each player is given $2 \ell$ cards, $\ell$ \mybox{$\clubsuit$}s and $\ell$ \mybox{$\heartsuit$}s. An $i$-th player has an integer $a_i \in \mathbb{Z}/k\mathbb{Z}$, indicating the candidate he/she prefers. We write $a_i$ in a binary representation $a_i = \sum_{j=0}^{\ell-1} 2^ja_{(i,j)}$, where $a_{(i,j)} \in \{0,1\}$ for every $j=0,1,...,\ell-1$. The $i$-th player then uses two cards to form a commitment of a bit $a_{(i,j)}$ for each $j=0,1,...,\ell-1$.

We have $E(a_1,a_2,...,a_n) = 1$ if and only if $E(a_{(1,j)},a_{(2,j)},...,a_{(n,j)}) = 1$ for every $j=0,1,...,\ell-1$. Therefore, we can use our second protocol to compute $b_j = E(a_{(1,j)},a_{(2,j)},...,a_{(n,j)})$ for each $j=0,1,...,\ell-1$. Then, we have $E(a_1,a_2,...,a_n) = b_0 \wedge b_1 \wedge ... \wedge b_{\ell-1}$ as desired.

We use $n-1$ shuffles to compute each bit $b_j$, and $\ell-1$ shuffles to compute the AND function of $\ell$ bits. Therefore, this protocol uses $2\ell n = 2 \lceil \lg k \rceil n$ cards and $\ell(n-1)+\ell-1 = \ell n-1 = \lceil \lg k \rceil n-1$ shuffles, and it is also committed-format.

\subsection{Comparison with Other Protocols}
As we use $\ell$ bits to represent each integer $a_i$, the $k$-candidate $n$-variable equality function can be viewed as a Boolean $\ell n$-variable function. Hence, we can directly apply the protocol of Nishida et al. \cite{nishida} to compute this function (which is not symmetric), which requires $2\ell n+6 = 2\lceil \lg k \rceil n+6$ cards and as much as $O(\lceil \lg k \rceil n \cdot 2^{\lceil \lg k \rceil n}) = O(\lg k \cdot n(2k)^n)$ shuffles. Therefore, our protocol uses fewer number of cards and much fewer number of shuffles than their protocol.

\section{Future Work}
For computing the Boolean equality function, our second protocol is already optimal in terms of number of cards as it matches a trivial lower bound of $2n$, and it is also committed-format. For computing other doubly symmetric functions, our first protocol is optimal in terms of number of cards but is not committed-format. This leaves an open problem to find a committed-format protocol to compute such doubly symmetric functions using $2n$ cards.

For computing symmetric functions that are not doubly symmetric, there is still a gap between our $2n+2$ and the trivial lower bound of $2n$ in terms of number of cards. This leaves an open problem to find a protocol that requires less than $2n+2$ cards, or to prove the lower bound of the number of required cards, both for a committed-format protocol and for any protocol.

\appendix
\section{Security of Protocols} \label{app1}
In some steps of our protocol, we turn a set of cards face-up. In this appendix, we will formally prove that seeing the front side of those cards does not leak any information.

\subsection{Addition Protocol} \label{app1.1}
In the addition protocol, after we shift $Y$ to the left by $r$ positions for a uniformly random $r \in \{0,1,...,k-1\}$, we turn over all cards in $Y$ and to reveal the sequence $E_3^\clubsuit(b+r\text{ (mod }k))$. Note that $b+r$ (mod $k$) has an equal probability of $1/k$ to be each integer in $\{0,1,...,k-1\}$ no matter what $b$ is.

For each $i \in \{0,1,...,k-1\}$ and $V \in \{E_3^\clubsuit(0), E_3^\clubsuit(1), ..., E_3^\clubsuit(k-1)\}$, consider the conditional probability of $b=i$ given that the visible sequence is $V$.

From Bayes' theorem, we have
\begin{align*}
\Pr(b=i | \text{visible}=V) &= \frac{\Pr(\text{visible}=V | b=i )\Pr(b=i)}{\Pr(\text{visible}=V)} \\
&= \frac{(1/k)\Pr(b=i)}{1/k} \\
&= \Pr(b=i).
\end{align*}

Therefore, one cannot deduct any information about $b$ upon seeing the face-up cards.

\begin{figure}[H]
\centering
\begin{tikzpicture}[node distance=2.3cm, auto, scale=0.8, every node/.style={scale=0.8}]
    \node [block] (a1) {\makecell[l]{
    $\clubsuit$$\heartsuit$...$\heartsuit$ $X_0$ \\
    $\heartsuit$$\clubsuit$...$\heartsuit$ $X_1$ \\
    ... \\
    $\heartsuit$$\heartsuit$...$\clubsuit$ $X_{k-1}$
    }};
    \node [block, below of=a1] (a2) {\makecell[l]{
    $\clubsuit$$\heartsuit$...$\heartsuit$ $\frac{1}{k}X_0 + \frac{1}{k}X_1 + ... + \frac{1}{k}X_{k-1}$ \\
    $\heartsuit$$\clubsuit$...$\heartsuit$ $\frac{1}{k}X_0 + \frac{1}{k}X_1 + ... + \frac{1}{k}X_{k-1}$ \\
    ... \\
    $\heartsuit$$\heartsuit$...$\clubsuit$ $\frac{1}{k}X_0 + \frac{1}{k}X_1 + ... + \frac{1}{k}X_{k-1}$
    }};
    \path [line] (a1) -- node{(\texttt{left\_shift},$r$)} (a2);
\end{tikzpicture}

\begin{minipage}{\textwidth}
	\centering
	\caption{A KWH-tree of the steps right before we turn over all cards in $Y$}
	\label{tree1}
\end{minipage}
\end{figure}
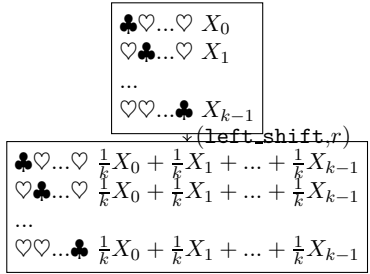

This proof can be illustrated by a KWH-tree, a tool developed by Koch et al. \cite{koch}. In the KWH-tree in Figure \ref{tree1}, each $X_i$ denotes the probability that $b=i$ (which is left as a variable as the distribution may be arbitrary), and a polynomial of variables denotes the probability that the sequence $Y$ is the one next to the polynomial.

\subsection{Revealing Commitments of $a'_n$} \label{app1.2}
In our first protocol and its generalization, after we XOR every input bit $a_i$ with a uniformly random bit $r \in \{0,1\}$, we turn over the two rightmost cards to reveal the commitments of $a_n \oplus r$. Note that $a_n \oplus r$ can be 0 or 1, each with probability 1/2, no matter what $a_n$ is.

For each $i \in \{0,1\}$ and $V \in \{$\mybox{$\clubsuit$}\mybox{$\heartsuit$}, \mybox{$\heartsuit$}\mybox{$\clubsuit$}$\}$, consider the conditional probability of $a_n=i$ given that the visible sequence is $V$.

From Bayes' theorem, we have
\begin{align*}
\Pr(a_n=i | \text{visible}=V) &= \frac{\Pr(\text{visible}=V | a_n=i )\Pr(a_n=i)}{\Pr(\text{visible}=V)} \\
&= \frac{(1/2)\Pr(a_n=i)}{1/2} \\
&= \Pr(a_n=i).
\end{align*}

Therefore, one cannot deduct any information about $a_n$ upon seeing the face-up cards.

\begin{figure}[H]
\centering
\begin{tikzpicture}[node distance=2.3cm, auto, scale=0.8, every node/.style={scale=0.8}]
    \node [block] (a1) {\makecell[l]{
    $\clubsuit$$\heartsuit$ $X_0$ \\
    $\heartsuit$$\clubsuit$ $X_1$
    }};
    \node [block, below of=a1] (a2) {\makecell[l]{
    $\clubsuit$$\heartsuit$ $\frac{1}{2}X_0+\frac{1}{2}X_1$ \\
    $\heartsuit$$\clubsuit$ $\frac{1}{2}X_0+\frac{1}{2}X_1$
    }};
    \path [line] (a1) -- node{random bit XOR operation} (a2);
\end{tikzpicture}

\begin{minipage}{\textwidth}
	\centering
	\caption{A KWH-tree of the steps right before we turn over the two rightmost cards}
	\label{tree2}
\end{minipage}
\end{figure}
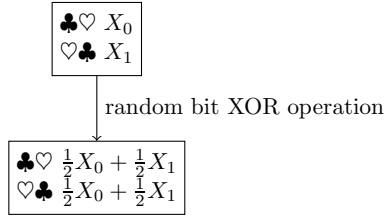

This proof can be illustrated by a KWH-tree in Figure \ref{tree2}, where each $X_i$ denotes the probability that $a_n=i$, and a polynomial of variables denotes the probability that two rightmost cards are the one next to the polynomial.

\subsection{Final Random Cuts} \label{app1.3}
In our first protocol and its generalization (in this proof we consider the generalization as it also covers the original protocol), let $s$ be the final sum of $n-1$ bits. After we apply the final random cuts for every $b \in \Ima f$ such that $|P_b|>1$, we turn over all cards in $Y$. Suppose the visible sequence is $E_n^\heartsuit(s')$ and thus the protocol returns $g(s')=b'$.

If $|P_{b'}|=1$, then we know that $s=s'$, which is inevitably a public information given that we know the output $b'$. If $|P_{b'}|>1$, the public information deducted from $b'$ is only that $s \in P_{b'}$. Because of the final random cuts, a sequence $E_n^\heartsuit(i)$ for each $i \in P_{b'}$ has an equal probability of $1/|P_{b'}|$ of appearance no matter what $s$ is.

For each $i \in P_{b'}$ and $V \in \{E_n^\heartsuit(j) | j \in P_{b'}\}$, consider the conditional probability of $s=i$ given that the visible sequence is $V$.

From Bayes' theorem, we have
\begin{align*}
\Pr(s=i | \text{visible}=V) &= \frac{\Pr(\text{visible}=V | s=i )\Pr(s=i)}{\Pr(\text{visible}=V)} \\
&= \frac{(1/|P_{b'}|)\Pr(s=i)}{1/|P_{b'}|} \\
&= \Pr(s=i).
\end{align*}

Therefore, one cannot deduct any information about the actual sum $s$ beyond the public information $s \in P_{b'}$ upon seeing the face-up cards.

\end{document}